\documentclass[letterpaper]{article} 
\usepackage{aaai24}  
\usepackage{times}  
\usepackage{helvet}  
\usepackage{courier}  
\usepackage[hyphens]{url}  
\usepackage{graphicx} 
\usepackage{natbib}  
\usepackage{caption} 
\usepackage{algorithm}
\usepackage{algorithmic}
\usepackage[hang,flushmargin]{footmisc}
\usepackage{colortbl}
\usepackage{xcolor}
\usepackage{subcaption}
\usepackage{utfsym}
\usepackage[export]{adjustbox}
\usepackage{amssymb}
\usepackage{bm}

\usepackage{newfloat}
\usepackage{listings}
\DeclareCaptionStyle{ruled}{labelfont=normalfont,labelsep=colon,strut=off} 
\floatstyle{ruled}
\newfloat{listing}{tb}{lst}{}
\floatname{listing}{Listing}
\usepackage{booktabs}
\usepackage{multirow, makecell}
\newcolumntype{L}[1]{>{\raggedright\let\newline\\\arraybackslash\hspace{0pt}}m{#1}}
\newcolumntype{C}[1]{>{\centering\let\newline\\\arraybackslash\hspace{0pt}}m{#1}}
\newcolumntype{R}[1]{>{\raggedleft\let\newline\\\arraybackslash\hspace{0pt}}m{#1}}

\title{Automating Transparency Mechanisms in the Judicial System\\ Using LLMs: Opportunities and Challenges} 

\author {
    Ishana Shastri\textsuperscript{\rm 1},
    Shomik Jain\textsuperscript{\rm 1},
    Barbara Engelhardt\textsuperscript{\rm 2},
    Ashia Wilson\textsuperscript{\rm 1}
}
\affiliations{
    \textsuperscript{\rm 1}Massachusetts Institute of Technology
    \textsuperscript{\rm 2}Stanford University\\[0.5mm]
    ishana@mit.edu, shomikj@mit.edu, bengelhardt@stanford.edu, ashia@mit.edu
}

\begin{document}
\maketitle

\begin{abstract}
Bringing more transparency to the judicial system for the purposes of increasing accountability often demands extensive effort from auditors who must meticulously sift through numerous disorganized legal case files to detect patterns of bias and errors. For example, the high-profile investigation into the Curtis Flowers case took seven reporters a full year to assemble evidence about the prosecutor's history of selecting racially biased juries. LLMs have the potential to automate and scale these transparency pipelines, especially given their demonstrated capabilities to extract information from unstructured documents. We discuss the opportunities and challenges of using LLMs to provide transparency in two important court processes: jury selection in criminal trials and housing eviction cases. 
\end{abstract}

\section{Introduction}
The criminal legal system is known to be structurally biased in ways that amplify existing patterns of social inequality~\cite{appleman2016nickel, hetey2018numbers, hinton2018unjust}. However, most evidence of this state of affairs comes from the diligent work of reporters and researchers, with particularly egregious cases receiving additional scrutiny~\cite{apmReport, cohen2019judicial}. More transparency is needed at scale and over time across all components of the criminal legal system. While such transparency may or may not uncover illegal actions, regular audits are known to improve adherence to standards by discouraging improper behavior through implied oversight~\cite {raji2019actionable}. Legal structures such as malpractice claims, the Department of Justice's Office for Professional Responsibility, and more recently, prosecutorial conduct committees, are in place to monitor these biases. But these avenues fail to impose consequences for biased behaviors, largely because they are unable to uncover and process the documents that detail biased conduct. In this work, we argue that a substantial opportunity exists for AI systems to automate and scale existing transparency mechanisms, which could allow lawyers and academics to more rapidly conduct audits and address injustices.

There are a number of challenges to automating parts of the existing data-driven transparency mechanisms. First, legal documents are difficult to access due to a lack of centralization and digitization. Clerk's offices hold onto printed casebooks until they reach their expiration date; while the information is available to the public, accessing court files often requires going to courthouses, pulling relevant files, and manually scanning them. A second concern is the difficulty in processing the varying types and formats of data the legal system collects. In criminal proceedings, most state courts require some kind of record keeping such as court transcripts, but documentation requirements vary from state to state, and the data is non-standard and incomplete in a variety of ways. Even within a jurisdiction, court data is largely unstructured and often poorly annotated and logged; building technologies robust enough to handle such high variance is a challenge. Even the digitization of court documents into machine-readable formats remains a technical problem, especially given that some documents contain handwritten information. 

A key example of this data-driven transparency is an audit of jury selection done by journalists working on the American Public Media (APM) Reports podcast ``In The Dark,'' who spent a year gathering and analyzing court records from 1992 through 2017 in the Fifth Circuit Court District of Mississippi~\cite{apmReport}. With this data, APM was able to uncover patterns of discrimination in a series of trials that culminated in a man's exoneration and the resignation of a district attorney~\citep{dougEvansResigns}. According to one of the lead reporters, it took a team of seven people almost a year to process the court documents into a dataset for analysis~\cite{apmReport}. The process consisted of showing up with a scanner to each courthouse, going through each courthouse's docket book and writing down the names of each trial, then pulling the relevant case files. The team then spent months extracting the relevant data about jury selection from the case files of every trial. Researchers who conducted a similar audit in North Carolina concluded that the difficulties in data accessibility and processing leave ``no vantage point from which one might see the whole of jury selection, rather than the selection of a single jury''~\citep{wright2018jury}. 

Large language models (LLMs) have the potential to automate and scale transparency mechanisms given their promising information extraction capabilities. LLMs have been shown to extract information from many kinds of documents, such as full-length novels~\citep{chang2023booookscore}, electronic health records~\citep{van2023clinical}, and financial documents~\citep{kim2024bloated}. A growing area of research explores using LLMs for processing legal documents as well. However, the majority of these works focus on automating information extraction tasks performed by lawyers, such as contract review~\citep{hendrycks2021cuad}, case summarization~\citep{ash2024translating}, and legal reasoning tasks~\citep{guha2024legalbench}. 

Our work is one of the first to explore using LLMs for transparency in two important court processes: jury selection in criminal trials and housing eviction cases. Specifically, we make the following contributions:

\begin{itemize}
    \item We highlight the opportunities for LLMs to automate unstructured document extraction tasks that constitute current transparency mechanisms in each domain.
    \item We evaluate LLM performance across some of these tasks and surface several challenges, such as the various capabilities needed and nuanced sources of error.  
    \item We spotlight the need for both legal and technical investments to make automated transparency and auditing mechanisms more feasible.  
\end{itemize}

\subsection{Related Work}

The incredible proliferation of data-driven technologies across the US
criminal legal system is well-documented~\cite{barabas2020beyond,wang2018carceral}. These technologies have mostly prioritized the risk management of crimes, and work from the algorithmic fairness community has largely focused on ensuring these risk management tools satisfy technocratic notions such as accuracy and fairness~\cite{berk2021fairness, brayne2017big, chouldechova2017fair,goel2021accuracy}. Our work aligns with those who call to reimagine how AI systems are used in legal contexts. Specifically, as several works have pointed out, a more substantive understanding of what it means for AI to benefit carceral contexts would go from ensuring the measurement of defendants’ pathologies and deficiencies are ``fair'' and ``accurate'' to serving decarceral ends. As Chelsea Barabas further describes:  
{\em ``an abolitionist re-imagining of AI in criminal law would require shifting away from measuring criminal behavior and towards understanding processes of criminalization, from supporting law and order towards increasing community safety and self-determination, and
from surveilling risky populations towards holding accountable state officials''}~\cite{barabas2020beyond}.

Only a few studies have explored using LLMs for accountability or transparency in the legal domain. \citet{chien2024generative} focus on the potential for LLMs to make legal processes and information more accessible to low-end consumers. Specifically, they developed a GPT-powered chatbot to extract information on Arizona's eviction rules and provide guidance to users on eviction forms and procedures. \citet{pereira2024inacia} investigate the ability of GPT-4 to streamline the processing of Brazilian audit cases. In a pipeline similar to ours, they start with raw case documents and attempt to determine the allegations made as well as the legal admissibility and plausibility of the case. While they hope to assist audit courts in speeding the processing of cases, we differ in our goal of providing transparency into court cases after their resolution. We contribute a novel application of LLMs for automating transparency mechanisms for how cases are adjudicated.

\section{Case Studies}
In this section, we provide case studies of two important court processes with a history of biased and exploitative practices. We describe how transparency mechanisms in each domain rely on the manual analysis of numerous unstructured court documents in order to highlight the opportunities for LLMs to potentially help audit for bias. 

\subsection{Jury Selection}

Jury selection plays a central role in ensuring the fair and impartial administration of justice in criminal trials. However, the process has come under scrutiny for its opacity and implicit biases in which jurors are selected. Transparency into jury selection requires the analysis of long court transcripts and handwritten jury strike sheets, which LLMs could potentially help process. 

\subsubsection{Background.}

In the US, the process of narrowing down a pool of potential jurors, or the \textit{venire}, to the final list differs slightly across different jurisdictions and courts. However, the core procedure centers around the process termed \textit{voir dire}. Initially, potential jurors are randomly sampled from a state's voter and motor vehicle registration lists. These potential jurors then proceed through the voir dire process, where they answer a verbal questionnaire administered by the judge or counsel that aims to determine if jurors are impartial and capable of sitting on the jury for that case. Attorneys from each side of the case may ask questions such as if any health issues could potentially stand in the way of serving on the jury, if the potential juror has any pre-existing ties to the criminal justice system, and if the potential juror holds any preconceived notions or biases about the case type. 

While voir dire is intended to remove jurors who could be biased regarding the case, the process leaves doubts as to why jurors are struck. In a criminal trial, either the prosecutor or the state defense can strike a juror \textit{for cause} or through \textit{peremptory strikes}. Strikes for cause require a legal basis for a juror's dismissal to be provided to the judge, such as knowing the defendant or an inability to effectively communicate in court. Alternatively, peremptory strikes allow each side to dismiss a certain number of jurors without reason unless specifically questioned by the court. 

\subsubsection{Exploitative Practices.}
The format of peremptory strikes creates an easy loophole for biased strikes and jury manipulation~\citep{fixingbatson, bennett2010unraveling}. In the landmark 1986 \textit{Batson v. Kentucky} case, the US Supreme Court ruled that peremptory strikes could not be used to exclude a potential juror solely on the basis of race, and later the ruling was extended to include gender and sexual orientation as well. But the burden falls on the opposing party to raise a \textit{Batson challenge}, which requires the suspected party to provide a race- and gender-neutral reason for why they struck the juror. Moreover, these challenges are often ruled unsuccessful by judges when prosecutors provide vague reasons such as low intelligence, which are actually ``race-neutral'' ways to strike people of color \citep{bennett2010unraveling, fixingbatson}. For example, a prosecutor in the American Public Media (APM) dataset gave the following reason for striking a black male from the jury: \textit{``He has an earring in his ear. I do not like to keep jurors, male jurors that wear earrings''}~\citep{apmReport}.




The infamous \textit{Curtis Flowers v. Mississippi} trials (2000-2010) are a canonical example of the extent to which bias in jury selection can completely upend a case outcome and thus an innocent person's life. Flowers was tried six times for murder, four of which resulted in convictions and the death sentence. These dispositions were later overturned by the Supreme Court, with the ruling citing a ``relentless, determined effort to rid the jury of black individuals'' by Doug Evans, the prosecutor in the cases~\citep{flowersSupremeCourt}. Flowers' case does not stand in isolation -- the Groveland Four in 1949 faced a similar outcome when two innocent black men received death sentences and one other was sentenced to life in prison for a crime they did not commit. Again, the all-white jury that the prosecutors selectively assembled resulted in a case outcome that only years later was overturned for racial bias. More recently, peremptory strikes against female and Jewish jurors in death row cases have come under question as instances of illegal bias, which could lead to more high-profile overturned convictions~\cite{female_juror, jewish_juror_california}.


\subsubsection{Transparency Mechanisms.}


Audits conducted by journalists and social scientists can help uncover specific prosecutors and jurisdictions with patterns of implicit bias in jury selection~\citep{apmReport, wright2018jury}. The aforementioned audit of Mississippi criminal trials conducted by American Public Media (APM) Reports found that prosecutors exercised a disproportionate number of peremptory strikes against black jurors, striking them at a rate 4.5 times that of white jurors. Female jurors were also struck by the prosecution at a rate 1.2 times that of male jurors~\citep{apmReport}. These findings helped pressure Doug Evans, the prosecutor in the Curtis Flowers' case, to resign over broader allegations of racial bias in cases spanning his 30 years as a Mississippi district attorney~\citep{dougEvansResigns}. APM's audit took a team of seven people a full year to compile an aggregated dataset of jurors, whether they were selected or struck, and demographic information over about 305 criminal trials from 1992 to 2017 in Mississippi's Fifth Circuit Court District. 

\subsubsection{Document Extraction Tasks.}

In APM's dataset of Mississippi criminal trials, the jury selection information for each case is present in either (1) a court transcript of the jury selection process or (2) a jury strike sheet\footnote{The raw case files are available at: \url{https://features.apmreports.org/in-the-dark/season-two/source-notes.html}}. Court transcripts include the voir dire questionnaire process for each juror, as well as the final jury roll call. Both sections of the transcript possibly reveal which jurors were chosen and their gender (based on identifiers like `Mr.' or `Mrs.' and other pronouns used to refer to them in the transcript). Jury strike sheets include a complete list of all summoned jurors as well as demarcations for who was struck or chosen\footnote{Figure 3 in the Appendix shows example strike sheets.}. Sometimes, the prosecutor for the case includes handwritten notes about why a juror was struck and their race and gender, usually coded as `W' or `B' for White/Black, and `M' or `F' for Male/Female. Other states besides Mississippi mandate the collection of demographic information from prospective jurors, which may or may not be included in the case files~\citep{jurySelectionNationwide}. Assuming the information is present, LLMs could potentially automate the following extraction tasks from court transcripts, jury strike sheets, and other case documents:

\begin{itemize}
    \item \textbf{Juror Demographic Information}: name, race, gender, and occupation history;
    \item \textbf{Trial Information}: county, judge, attorneys, offense, and case verdict;
    \item \textbf{Voir Dire Responses}: reasons jurors are unable to be impartial to a case (ties to or biases about the law enforcement system, inability to communicate, etc.);
    \item \textbf{Selected Jurors}: whether each prospective juror was selected to serve on the jury, selected as an alternate, or struck for cause;
    \item \textbf{Batson Challenges}: whether or not a challenge claim was made, and by which party (prosecution or defense).
\end{itemize}

\subsection{Eviction}

Evictions occur when tenants of a rental property are forced to leave by the landlord, often through a court-based process. However, this process may allow landlords to exploit their power imbalance, especially in lower-income or minority communities. Transparency into the eviction process requires the analysis of many different court documents that are often unorganized and contain handwritten information. 

\subsubsection{Background.} 

The eviction process varies across US cities and jurisdictions~\citep{evictionLabFAQ}. Most evictions happen because tenants fail to pay their rent on time. Other reasons for eviction may involve violations of the lease, damages to the property, or otherwise caused disturbances. Some cities also allow for \textit{no-fault evictions}, in which a landlord seeks to regain possession of the property without claiming any faults by the tenant. 

To start the eviction process, landlords are usually required to provide written notice to the tenant, commonly called a \textit{Notice to Quit} (NTQ). If the tenant does not vacate or meet the landlord's demands, the landlord may file a lawsuit. In these instances, tenants are served notice to appear in court, such as through a \textit{Summons and Complaint} (S\&C) form\footnote{Figure 4 in the Appendix shows an example S\&C form.}. Landlords receive a default judgment to evict if tenants do not appear in court. If a tenant meets the landlord in court, several outcomes could occur -- the case may be dismissed voluntarily or by the court, it could go to trial, or be resolved through a settlement~\citep{evictionLabFAQ}. A settlement could still involve a move-out agreement, or result in a move-out later if tenants do not meet certain conditions~\citep{nicolePaper}. If the court issues an execution, law enforcement can forcibly remove a tenant.

\subsubsection{Exploitative Practices.} 
Several studies have documented a variety of injustices in the eviction process, and being evicted can make it considerably more difficult to find future housing~\citep{desmond2015eviction}. Landlords may exploit their power imbalance or tenants' lack of legal knowledge in order to intimidate them during the eviction process. Because eviction cases usually occur in civil court, where tenants have no right to an attorney, many tenants do not appear in court and receive a default judgment if the landlord is present. In fact, around 70\% of tenants do not show up to court for eviction cases in many major US cities~\citep{desmond2015eviction}. 

Some of these cases may involve no-fault or retaliatory evictions. No-fault evictions have historically been used to displace low-income tenants in areas with rent control in order to gentrify cities~\citep{desmond2016whogetsevicted}. Landlords may also issue retaliatory evictions where they threaten to evict as a consequence of tenants complaining about poor living conditions. An exploitative landlord may also make living conditions so unbearable that a tenant voluntarily evicts themselves. Restriction of electricity or heating access, coercion, or creating a hostile environment fall under this type of constructive or ``self-help'' eviction. While these practices are usually illegal, the burden falls on the tenant to successfully prove this in court. This may be difficult for tenants who cannot afford counsel or who are otherwise unable to make a strong case for themselves.

\subsubsection{Transparency Mechanisms.}

Legal academics such as the Eviction Lab at Princeton University have been able to uncover exploitative practices and patterns of bias through vigorous data collection and processing efforts. The Eviction Lab has been at the forefront of transparency work in eviction thus far, combining public eviction records\footnote{Transparency into evictions can also require documents beyond court files, such as tenant surveys to uncover informal evictions or mortgage records to trace connections between landlords under the same ownership.} with census data and proprietary individual records to assemble the largest dataset covering all 50 US states~\cite{elabreport}. Their work has helped to reveal how tenants with children, with low incomes, or in disadvantaged neighborhoods are disproportionately affected by evictions~\citep{desmond2016whogetsevicted}. Moreover, their case study of Milwaukee's inner-city neighborhoods found black renters were twice as likely to be evicted through the courts than white renters, and female renters were more than twice as likely to be evicted as male renters~\cite{desmond2012eviction,desmond2015eviction}.

Another academic effort to provide transparency into evictions is the recent work by \citet{nicolePaper}. They curated a dataset of eviction cases in Massachusetts through the meticulous process of physically accessing eviction files in the courthouse, scanning them, and then hand-coding them for analysis. Their work helps bring transparency into the various legal procedural pathways that result in forced tenant moves once an eviction case is filed. In particular, they find that move-out agreements are a settlement type often omitted from administrative datasets, yet are a primary procedural pathway by which tenants are forcibly moved~\citep{nicolePaper}.

\subsubsection{Document Extraction Tasks.}

We summarize the information extraction tasks undertaken by \citeauthor{nicolePaper} in order to explore whether LLMs could partially automate their extensive effort. The case files that they collected included the Notice To Quit (NTQ), Summons and Complaint (S\&C), counterclaims from the tenant, court order information, and other court documents. These files were often not arranged in chronological order and needed to be processed together in order to glean the following information:

\begin{itemize}
    \item \textbf{Case Background}: address, whether the tenancy was subsidized, the type of landlord, whether each party was represented by legal counsel, the type of eviction case (nonpayment of rent, fault, or no-fault); 
    \item \textbf{Procedural History of the Case}: whether the tenant defaulted, whether execution issued, case dispositions;
    \item \textbf{Settlement Terms}: specific settlement conditions, whether a judgment was entered in favor of the landlord.
\end{itemize}

\section{Methods for LLM Experiments}
We first discuss the various capabilities that LLMs would need in order to perform the document extraction tasks in our case studies. We then describe the subset of tasks that we test in our experiments for each domain. Our selected tasks are not meant to demonstrate an end-to-end automated pipeline. Rather, we explore tasks that we hypothesized to be feasible using LLMs and that test for different capabilities. Table~\ref{table:task_complexity} outlines our selected tasks and the capabilities required for each task.

\begin{table*}[ht]
\small
\centering
\begin{tabular}{L{4cm}L{3cm}C{1.45cm}C{1.45cm}C{2.25cm}C{2cm}}

\toprule
\addlinespace
\textbf{Task} & \textbf{Input Document} & \textbf{Synthesis} & \textbf{Inference} & \textbf{Non-Categorical Query} & \textbf{Handwritten Information} \\
\addlinespace
\toprule
\addlinespace
Selected Juror Names & \multirow{3}{3cm}{Court Transcript} & $\usym{2613}$ & $\usym{2613}$ & $\usym{1F5F8}$ & $\usym{2613}$ \\
Batson Challenges &  & $\usym{2613}$ & $\usym{1F5F8}$ & $\usym{2613}$ & $\usym{2613}$ \\
Jury Gender Composition &  & $\usym{1F5F8}$ & $\usym{1F5F8}$ & $\usym{1F5F8}$ & $\usym{2613}$ \\
\addlinespace
\midrule
\addlinespace
Zip Code & \multirow{3}{3cm}{Summons \& Complaint} & $\usym{2613}$ & $\usym{2613}$ & $\usym{1F5F8}$ & $\usym{2613}$ \\
Landlord Type & & $\usym{2613}$ & $\usym{1F5F8}$ & $\usym{2613}$ & $\usym{2613}$ \\
Landlord Representation Status & & $\usym{2613}$  & $\usym{1F5F8}$ & $\usym{2613}$ & \bm{$\thicksim$} \\
\addlinespace
Settlement Type & \multirow{3}{3cm}{Various Case Files} & $\usym{2613}$ & $\usym{1F5F8}$ & $\usym{2613}$ & \bm{$\thicksim$} \\
Case Disposition &  & $\usym{1F5F8}$ & $\usym{1F5F8}$ & $\usym{2613}$ & \bm{$\thicksim$} \\
Execution Issued & & $\usym{1F5F8}$ & $\usym{1F5F8}$ & $\usym{2613}$ & \bm{$\thicksim$} \\

\addlinespace
\bottomrule
\addlinespace
\multicolumn{6}{c}{$\usym{1F5F8}$ = capability always required, \bm{$\thicksim$} = capability required for some cases, $\usym{2613}$ = capability not required}
\end{tabular}
\caption{Document extraction tasks that we test in our experiments and the corresponding capabilities required to complete them (based on the court cases in the APM jury selection dataset and the \citeauthor{nicolePaper} eviction dataset).}
\label{table:task_complexity}
\end{table*}

\subsection{LLM Capabilities}Given how relevant information is presented in documents and the query format, the following capabilities are required for extraction. 
\begin{itemize}
    \item \textbf{Synthesis}: The relevant information is present in multiple sections of a document or across multiple documents.
    \item \textbf{Inference}: Answering the query requires logical or legal inference from the information in the document(s).
    \item \textbf{Non-Categorical Query}: The query does not ask for specific categorical outputs.
    \item \textbf{Handwritten Information}: The relevant information is handwritten as annotations in the document.
\end{itemize}
We focus on the case where LLMs are given machine-encoded text from documents that have already been converted using an external Optical Character Recognition (OCR) tool, such as Adobe Acrobat or Microsoft Azure. Thus, the LLM capability to process handwritten information refers to understanding text that may not have been converted perfectly by virtue of it being interspersed with the printed text or along the margins of the document.

\subsection{Jury Selection Experiments}

Using APM's dataset of Mississippi criminal trials~\citep{apmReport}, we focus on the feasibility of automating information extraction from court transcripts. We chose this because court transcripts have the potential to increase transparency even beyond jury selection, and because we found OCR limitations when processing the jury strike sheets. However, we still limit our analysis to the 50 cases in APM's dataset that had a strike sheet in addition to a final jury roll call in the transcript (see Figure~\ref{fig:example-transcript} for an example of the roll call). Focusing on these cases allowed us to cross-reference the anonymized APM dataset of jurors with the information in the strike sheet, as well as test the effect of fine-tuning on the portion of the transcript with the final jury roll call. We used Adobe Acrobat's OCR technology to extract the text from the scanned transcripts, which ranged anywhere from 15 pages to 400 pages.

We test the following tasks which are related to those performed by APM (c.f. Case Studies). However, we structure the juror name and gender composition tasks so that it is possible to determine them using only the final jury roll call. 

\begin{itemize}
    \item \textbf{Selected Juror Names}: the names of jurors selected to serve on the jury (including alternates).
    \item \textbf{Batson Challenges}: whether or not a challenge claim was made, and by which party (prosecution or defense).
    \item \textbf{Jury Gender Composition}: the counts of male and female jurors selected for the jury (including alternates).
\end{itemize}
Extracting the names of the final venire of jurors requires no synthesis or inference because they are always present at the end of the voir dire transcript as part of a final roll call by the judge. Determining whether a Batson challenge occurred during a case and from which party (prosecution or defense) requires the LLM to infer what constitutes a Batson violation and track which party initiated it. Determining the jury's gender composition requires both synthesis and inference by knowing which jurors were selected, finding their gender (specified through pronouns or prefixes like ``Mr.'' or ``Mrs.''), and outputting the final count of female and male jurors. 

\subsection{Eviction Experiments}

Using \citeauthor{nicolePaper}'s dataset of eviction cases in Massachusetts, we explore the feasibility of automating a few of their information extraction tasks. We limit our analysis to case books from 2013 that include the aforementioned NTQ, S\&C, and other court documents (105 cases). We generated an abridged version of each case book, removing miscellaneous notices, email records, and other pages not functional for our selected tasks. We used Microsoft's Azure OCR model\footnote{We originally tried Adobe Acrobat OCR on these documents as well, but found Microsoft Azure Document Intelligence to have better performance.} to extract both the text and the handwritten components from the scanned documents.

We test the following tasks that represent some of the variables (or aggregations of variables) coded by \citeauthor{nicolePaper}. For a complete list of variables coded by \citeauthor{nicolePaper} and their legal definitions, we refer to Appendix A in their cited work. We note that automating the extraction of zip code and landlord type would not provide additional transparency as these are often available in administrative datasets; however, we include them as a baseline for the other tasks.
\begin{itemize}
    \item \textbf{Zip Code}: zip code of the property.
    \item \textbf{Landlord Type}: whether the listed plaintiff is a ``Corporation'', ``Individual'', or the ``Boston Housing Authority'' (for public housing cases).
    \item \textbf{Landlord Representation Status}: whether the landlord was represented by legal counsel, determined by if an attorney signed the S\&C\footnote{Signatures on other forms would need to be checked if the landlord received limited assistance representation, but this did not occur in any of the considered cases.} on behalf of the landlord.
    \item \textbf{Settlement Type}: whether there was a move-out agreement, civil probationary agreement, or no settlement.
    \item \textbf{Ultimate Case Disposition}: voluntary dismissal, default judgment, dismissed by court order, trial by judge, or settlement agreement.
    \item \textbf{Execution Issued}: whether or not an execution to evict the tenant was issued by the court (and not later dismissed).
\end{itemize}
Determining the zip code requires no synthesis or inference as it is always present in the address section of the S\&C. The landlord type requires inference as either the landlord's name or the corporation's name is listed on the S\&C. Determining the landlord's representation status requires inference from signatory names on the S\&C, which may be handwritten. The settlement type also requires inference from whether there is a clause to vacate the property by a certain date, or if there is a clause for reinstatement of tenancy in accordance with specific terms. Determining the ultimate case disposition and whether an execution was issued both require parsing through multiple different files within the case book -- files such as a ``Notice of Dismissal'' or a settlement ``Agreement for Judgment'' are often included, indicating the type of resolution the case had. However, sometimes these files are not included in the casebook itself, but the disposition is instead handwritten on the case docket sheet\footnote{Figure 5 in the Appendix shows an example docket sheet.}. Additionally, cases may go through several different dispositions over time, making it necessary for an LLM to be able to synthesize across a chronologically ordered collection of documents. Detecting an execution requires understanding the different execution pipelines and distinguishing between motions to issue an execution that were allowed, withdrawn, or denied.

\subsection{Model and Experimental Details}
For our main analysis, we use OpenAI's GPT-4 Turbo model (\texttt{gpt-4-turbo-2024-04-09}). We use \texttt{gpt-3.5-turbo-0125} for our fine-tuning experiments. We evaluate GPT models because prior work found them to have the best performance in the legal domain~\citep{guha2024legalbench, pereira2024inacia}. Our main analysis also uses a zero-shot prompt structure. Table~\ref{tab:prompts} specifies the exact prompt that we use for each task, which we arrived at after testing variations on a few documents. For each court case and task, we conduct five iterations to account for stochasticity.

\section{Results and Challenges}
We find LLMs to have varying performance across tasks in both domains, and that accuracy is generally worse for tasks that require more capabilities. Table~\ref{tab:accuracy} reports the accuracy for each task using zero-shot prompts. The best performance is for the tasks that do not require synthesis or inference (juror names and zip code), which is expected as these tasks merely demand a simple search and return scheme. However, performance varies greatly across the more complex tasks. Below, we describe common errors and highlight task-specific nuances beyond capabilities that cause LLMs to struggle.

\begin{table*}[t!]
\small
\centering
\begin{tabular}{c|c|c|c|c} 
\toprule
Domain & \# Cases & Input Document & Task & Accuracy \\
\midrule
\multirow{4}{1.5cm}{\centering Jury\\Selection} & \multirow{4}{1cm}{\centering 50} & \multirow{4}{1.5cm}{\centering Full Transcript} & Selected Juror Names & 81.6 $\pm$ 4.8 \\
\cmidrule{4-5}
& & & Batson Challenges & 23.2 $\pm$ 5.2 \\
\cmidrule{4-5}
& & & Jury Gender Composition & 3.6 $\pm$ 2.3  \\
\midrule
\multirow{8}{1.5cm}{\centering Eviction} & \multirow{8}{1cm}{\centering 105} & \multirow{4}{1.5cm}{\centering Summons \& Complaint} & Zip Code & 95.8 $\pm$ 1.7  \\
\cmidrule{4-5}
& & & Landlord Type & 89.7 $\pm$ 2.6  \\ 
\cmidrule{4-5}
& & & Landlord Representation Status & 71.0 $\pm$ 3.9  \\
\cmidrule{3-5}
& & \multirow{4}{1.5cm}{\centering Various Case Files} &  Settlement Type & 88.6 $\pm$ 2.7 \\
\cmidrule{4-5}
& & & Case Disposition & 94.9 $\pm$ 1.9  \\ 
\cmidrule{4-5}
& & &  Execution Issued & 68.8 $\pm$  4.0\\
\bottomrule
\end{tabular}
\caption{Accuracy with 95\% CI by domain, input type, and task. Computed over all iterations (5 per case).}
\label{tab:accuracy}
\end{table*}

\subsection{Results for Jury Selection Tasks} 

\paragraph{Selected Juror Names (81.6\% accuracy).} We find two common errors that occur in the cases with an incorrect output. The first is incomplete recall, where the model ``forgets'' part of the answer to the task (i.e., some of the names are missing). The other error is misunderstanding the output format, such as outputting the juror IDs rather than their names. 

\paragraph{Batson Challenges (23.2\% accuracy).} The legal inference required for this task may contribute to the low accuracy. In particular, the LLM must know what indicates a successful Batson challenge and how to distinguish which party in the court initiated it. We experiment with two-shot prompting, as we elaborate on below, to facilitate in-context learning for the legal inference required. There were no observable patterns of errors for this task.

\begin{figure} [t!]
    \centering
    \captionsetup{width=\linewidth}
    \includegraphics[width=\columnwidth]{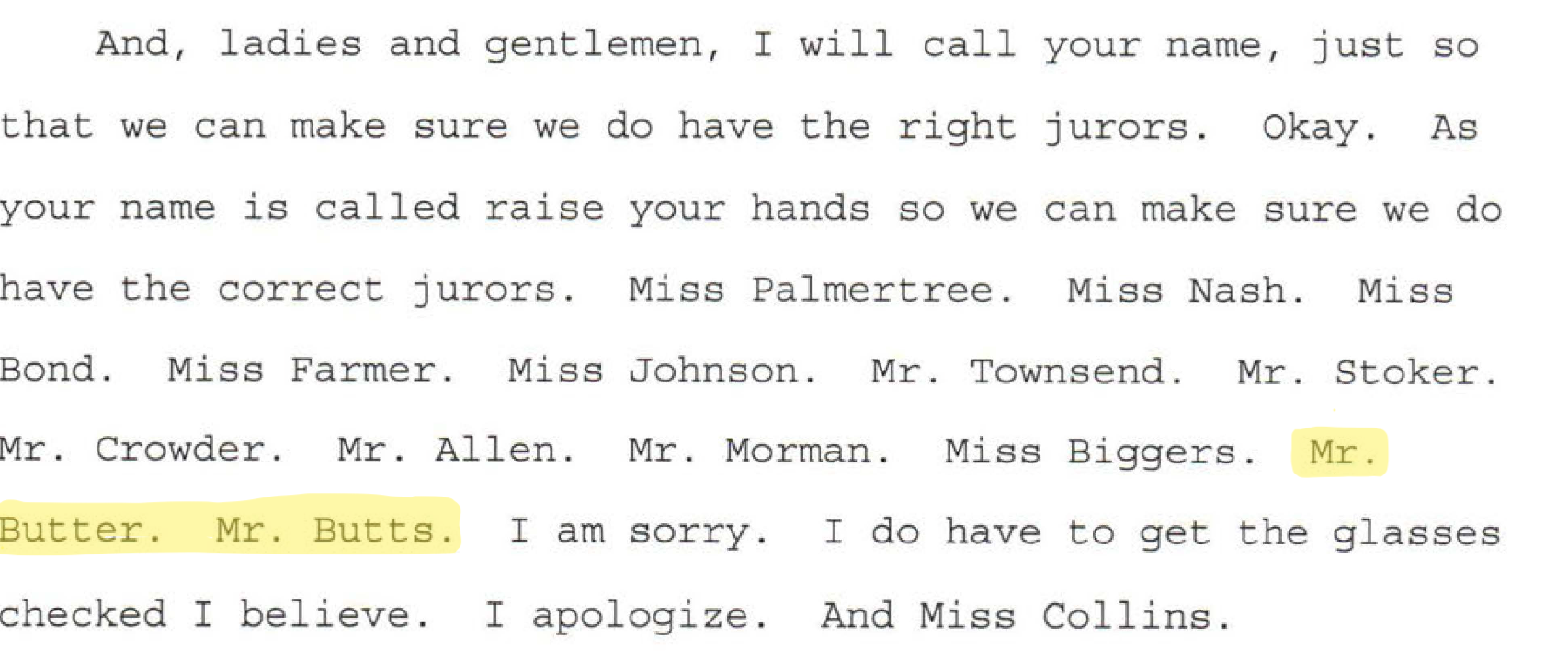}
    \caption{Example of a jury selection voir dire transcript excerpt. We extracted these excerpts of the final jury roll call in order to improve performance on the tasks of extracting selected juror names and determining the jury's gender composition. The highlighted text is a disfluency that causes the model to miscount jurors.}
    \label{fig:example-transcript}
\end{figure}

\paragraph{Jury Gender Composition (3.6\% accuracy).} This task had the worst accuracy among jury selection tasks. We hypothesize that one challenge is the synthesis required across the entire court transcript, as compared to the other jury selection tasks that can be isolated to specific sections of the transcript. We also observe a more complex failure point for this task related to understanding transcribed speech. For example, sometimes a prosecutor will misread a name and proceed to correct it, or call the same person twice (e.g., Figure \ref{fig:example-transcript}). These \textit {disfluencies}, or breaks and disruptions that occur in the flow of speech, may cause LLMs to misunderstand how many male or female jurors there actually are. We also note that the performance of gender composition is uncorrelated to the performance of extracting juror names, suggesting that the error may stem from complications in the aggregation step. 

\subsection{Results for Eviction Tasks}

\paragraph{Zipcode (95.8\% accuracy), Landlord Type (89.7\% accuracy), and Representation Status (71.0\% accuracy).} For these tasks, which require only the S\&C form, we observe that the performance decreases as the number of capabilities required increases. However, the synthesis and inference requirements do not reduce performance as much as in the jury selection tasks. One likely reason is that the S\&C form is shorter and more structured than a full court transcript. For zip code, the primary source of error is misunderstanding the prompt, such as outputting the zip code of the landlord's office instead of the property. The main type of error for landlord type and representation status was the model failing to find any relevant information. 

\paragraph{Case Disposition (94.9\% accuracy).} LLMs perform surprisingly well in determining the ultimate case disposition. This is likely because there are usually specific files corresponding to the ultimate disposition (e.g., a ``Notice of Dismissal'' for a dismissed case). A majority of errors come from cases that do not have these files present and thus require using the case docket sheet for the final disposition, where the information may be handwritten. 

\paragraph{Settlement Type (88.6\% accuracy).} While this task does not require synthesis, the performance is slightly worse than the case disposition task, which requires both synthesis and inference. We hypothesize that this is because determining the settlement type has a higher dependence on handwritten information than the case disposition in our dataset. The settlement type almost always has a handwritten component within the typed settlement agreement form, with cross-outs and check boxes for additional terms written by the plaintiff  (e.g., a common format is: $\square$ The tenant agrees to vacate on \underline{\hspace{0.6cm}}). We observe worse performance for settlement agreements with messier handwritten components, suggesting that the errors from the OCR process may propagate to the model's inference abilities. We also see errors stemming from the model failing to classify the case as a settlement case and thus not extracting any settlement terms.

\paragraph{Execution Issued (68.8\% accuracy).} This task had the lowest accuracy across all eviction tasks, which we attribute to several sources. First, determining whether an execution was issued requires substantially more legal context than other tasks, since there are several different pathways to an execution based on the case disposition~\cite{nicolePaper}. An execution to evict the tenant may be issued after (1) a default judgment, (2) a move-out settlement agreement, (3) if the tenant violates the terms of a civil probationary agreement and the landlord proceeds with the execution process, or (4) if there is a judgment for the landlord post-trial. Without a concrete understanding of these legal pathways, an LLM would be unable to correctly assess if a valid execution was issued to force the tenant to move out. Another source of difficulty is that the result of the execution motion (allowed, withdrawn, or declined) is often handwritten in the margins of the motion or shorthand on the docket sheet. Without appropriate OCR and inference capabilities to discern the motion result, the model can struggle to accurately categorize the case. Finally, there may be executions issued in cases that are ultimately dismissed; therefore, the LLM needs to understand the order of dispositions in order to determine whether the tenant was ultimately required to move out.

\begin{table*}[t!]
\small
\centering
\begin{tabular}{c|cccc} 
\toprule
Prompt & Zero-Shot & Two-Shot & Zero-Shot & Zero-Shot \\
Input Type & Full Transcript & Full Transcript & Roll Call Excerpt & Roll Call Excerpt \\
Fine-Tuning & No & No & No & Yes \\ 
\toprule
Selected Juror Names & 81.6 $\pm$ 4.8 & \textbf{94.8 $\pm$ 2.8} & -- & -- \\
\cmidrule{1-5}
Jury Gender Composition & 3.6 $\pm$ 2.3 &  18.4 $\pm$ 4.8  & 23.8 $\pm$ 5.3 & \textbf{34.0 $\pm$ 6.6} \\
\cmidrule{1-5}
Batson Challenges & 23.2 $\pm$ 5.2 &\textbf{ 76.8 $\pm$ 5.2} & -- & -- \\
\bottomrule
\end{tabular} 
\caption{Accuracy for jury selection tasks across experiment types. 95\% CI computed over all iterations (5 per case).}
\label{table:jury-improvements}
\end{table*}

\subsection{Improving Jury Selection Performance}

We explore few-shot prompting, reducing the document length, and fine-tuning as avenues to improve performance on jury selection tasks. In particular, we experiment with two-shot prompting for all three jury tasks but focus the latter two avenues on only the jury gender composition task because it had the worst performance across all tasks across both domains. Table~\ref{table:jury-improvements} reports the accuracy improvements for all tasks. Figure~\ref{fig:absolute_error} shows the improvements in absolute error, which we define as the sum of the differences in predicted and actual counts of male and female jurors, as well as the standard error for the gender composition task. Using the full transcript with zero-shot prompting, the baseline absolute error on average is 4.09 $\pm$ 0.305 (standard error).

\subsubsection{Few-Shot Prompting.}

For all three tasks, we find that accuracy improves with two-shot prompts compared to zero-shot prompts. In particular, we observe that the Batson challenges task has the largest jump in performance, from 23.2\% to 76.8\% accuracy. We also see a reduction in the absolute error for the gender composition task to 2.61 $\pm$ 0.263. This suggests that the legal context and inference capabilities for some tasks might be possible to inject into the model via few-shot prompting. 

\subsubsection{Document Length.}

Instead of using the full court transcripts, we test the jury gender composition task using just the excerpts of the final jury roll call to isolate the relevant information (e.g., Figure~\ref{fig:example-transcript}). This leads to an improvement in accuracy from 3.6\% with zero-shot prompts to 23.8\%. Moreover, if we examine the absolute errors, we observe both reduced error and variance across the documents (Figure \ref{fig:absolute_error}). This results in an improvement of the absolute error to 2.17 $\pm$ 0.233, similar to the performance of the two-shot prompting. While shortening the document length reduces the synthesis required, it does not address the problems of inferring gender and understanding disfluencies in transcribed speech.

\subsubsection{Fine-Tuning.}

\begin{figure}[t!] 
    \centering
    \includegraphics[width=0.85\columnwidth]{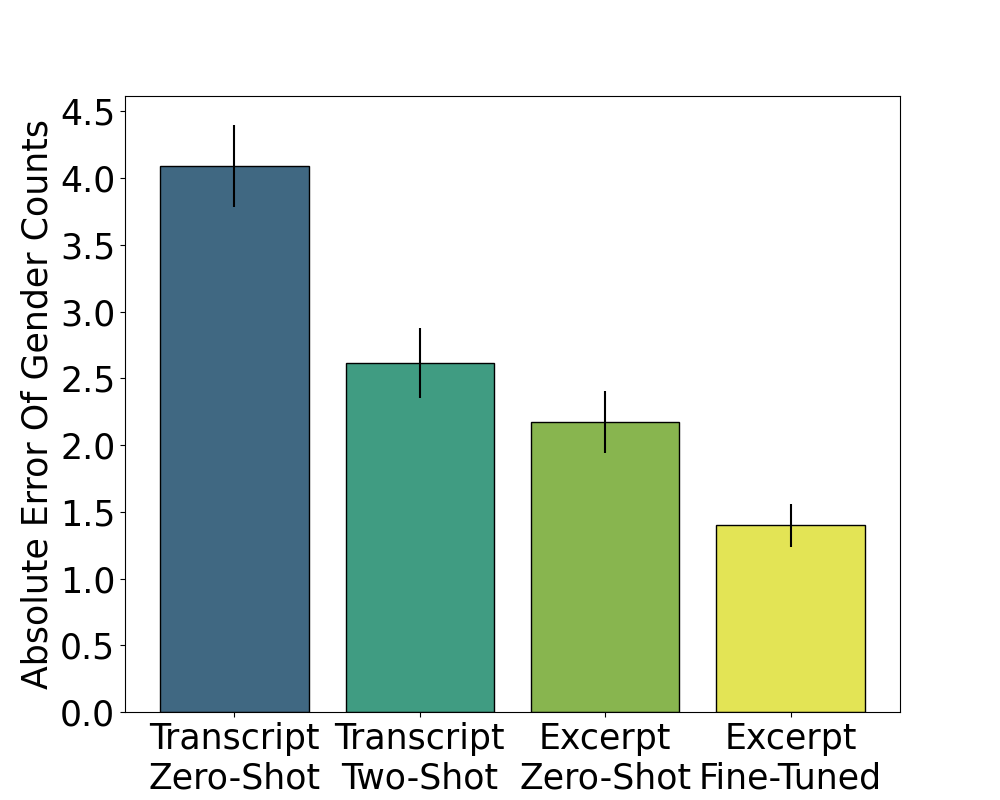}
    \caption{Absolute error for the jury gender composition task across different technical interventions. Error bars represent the standard error over all iterations.}
    \label{fig:absolute_error}
\end{figure}

We test if fine-tuning can help models understand disfluencies and improve performance on the gender composition task. We fine-tune on the excerpts of the final jury roll call with a 60-40 train-test split\footnote{In our sample of 50 cases, this corresponds to 30 cases in the training set and 20 cases in the test set.}, and average the results over ten random train-test splits. Fine-tuning using the excerpts only increases the accuracy from 23.8\% to 34.0\%. However, the absolute error and standard error on average were reduced to 1.40 $\pm$ 0.053, yielding an error reduction of 65.8\% and a standard error reduction of 82.6\% from the baseline (Figure \ref{fig:absolute_error}).

\subsection{Downstream Impact Tests}

Lastly, we highlight the importance of measuring model performance as it relates to downstream auditing questions that researchers and journalists might ask. We use the jury gender composition task as our working example. One goal of APM's analysis was to understand the selection ratios of different demographics on the jury and how they differed across specific counties and prosecutors~\citep{apmReport}. We analyze these downstream questions using our LLM outputs to see the extent to which model errors impacted the broader conclusions one could draw about the dataset. Specifically, we simulate what APM's analysis would look like using outputs from both our baseline LLM (using the full transcript as input) and fine-tuned LLM (using the roll call excerpt as input).

We find that using LLM outputs to answer downstream questions substantially alters the outcomes of a potential audit, although the fine-tuned model has less of an impact. The ground truth female-to-male ratio was 1.36 on average. This would decrease to 1.22 under the baseline LLM outputs, and 1.29 under the fine-tuned LLM outputs. However, both the baseline LLM and fine-tuned LLM would have flipped the dominant gender in the jury for 11 out of our 50 cases. We also find large changes in the ranking of counties and prosecutors with the most female bias in jury selection. Using the baseline model, the ranking changed for four of the seven counties and nine out of the ten prosecutors in the dataset (Tables 8 \& 9 in the Appendix). With the fine-tuned model, the ranking changed for only two of the seven counties and five of the ten prosecutors. These effects could be further exacerbated if tested with more jurisdictions and prosecutors, making it difficult to conclude that LLMs could be relied on to automate transparency in this scenario.

\section{Discussion}
Our results indicate the challenges of using LLMs to automate information extraction tasks that would provide more transparency in court processes. In this section, we discuss the technical and legal investments that are needed in order to make LLMs more feasible for legal auditing. 

\subsection{Technical Investments}

\paragraph{Re-Orienting Benchmarks.} Our results provide several examples of how LLM capabilities may not always translate to tasks with real-world impact. For example, our jury gender composition query highlights the difficulty of a simple counting task -- a synthesis capability for which models demonstrate high scores on toy datasets~\cite{mathchat2024}. Moreover, capabilities may not be robust to nuances across all tasks that concern a particular set of legal documents. An example of this is determining whether an execution was issued from eviction case files, which requires more layered deductions and familiarity with specific contexts than the other eviction tasks we tested. These results indicate the need to orient more LLM benchmarks around tasks with real-world impact, such as the applications we highlight. 

\paragraph{Training Datasets.} More training on the types of unstructured data in the legal domain could significantly improve performance for judicial transparency use cases. The poor out-of-the-box performance that we observe for court documents may be because these documents are not often digitized, and LLMs are mostly trained on easily scrapable internet datasets~\citep{llmInternet}. For example, transcribed speech may not be a common data format across online data, which would explain why LLMs struggle to understand speech disfluencies.

\paragraph{Pre-Processing Capabilities.}
Processing handwritten information is a major bottleneck for information extraction from legal documents. Important information is often handwritten by lawyers and judges in the margins of documents that are already unstructured. Even state-of-the-art optical character recognition (OCR) tools struggle with the high variance of messy handwriting in court documents. Another pre-processing task is to identify the relevant parts of documents and casebooks, as we found better performance with shorter input lengths. Methods such as Retrieval Augmented Generation (RAG) may be useful for this task, which could improve performance on longer casebooks with extraneous documents and forms. 

\subsection{Legal Investments}

\paragraph{Data Accessibility and Standardization.} States and jurisdictions could mandate more standard formats for documentation in court cases, and collect documents in digital databases. As we discuss in our case studies, accessing court files often requires going to courthouses and manually scanning files. Judges or clerks could also be required to convert handwritten notes into typed records. The potential benefits of LLMs cannot be fully realized unless documents are digitized and in more standard formats. Fine-tuning LLMs also requires a collection of labeled examples across various transparency-related tasks. The investment by law firms into collecting data for the types of tasks paralegals have traditionally performed (e.g., case summarization) may be helpful, but even these tasks may not directly translate into transparency use cases.

\paragraph{Model End-Users.} Journalists and legal practitioners who currently perform transparency mechanisms may be reluctant to adopt LLMs in their pipelines. LLMs may already be able to assist with easier tasks that have less downstream impact. However, for more complicated tasks, human experts may need to perform the overhead of prompt engineering and monitoring for errors. Some tasks may also not be possible to fully automate given the domain knowledge required, such as figuring out if a property tends to offer subsidized housing. Tasks like these would require human experts to oversee a model in order to check edge cases. Technical scholars must continue to collaborate with social scientists and legal practitioners in order to understand and address the hesitations in using LLMs to assist their work.

\paragraph{Mitigating Disparate Impacts.}
We caution against the potential disparate impacts of deploying LLMs for legal transparency without further investigation into the failure points and direct implications. Jurisdictions representing affluent neighborhoods and higher-profile cases generally have better-maintained documentation and could benefit more from the adoption of LLMs. Private lawyers and law enforcement agencies may also have the resources to keep better documentation than public defenders. Given that LLMs are likely to perform better on well-maintained and structured data, advantaged communities may be more likely to benefit from automated transparency pipelines, while minorities and those who would benefit the most from more transparency still face higher barriers.\\

We call on both the technical and legal communities to invest in solutions that can bring more transparency to the judicial system. By focusing efforts on legal data centralization, training models on unstructured data, and remaining vigilant about the implications of errors, information extraction using LLMs can help support efforts to provide transparency in the judicial system.

\section{Acknowledgments}
We are very grateful to Nicole Summers for sharing her dataset and providing us with feedback. We would also like to thank Will Craft, Matthew Desmond, and Carl Gershenson for their advice and insights. This research was funded in part by the Simons Collaboration on The Theory of Algorithmic Fairness and Lister Brothers Career Development Professorship at MIT. BEE is a CIFAR Fellow in the Multiscale Human Program and is on the Scientific Advisory Board for ArrePath Inc, Crayon Bio, and Freenome; she consults for Neumora.

\section{Appendix}
Supplementary materials are available on arXiv.

\begin{table*}[h!]
\small
\centering
\begin{tabular}{p{0.2\linewidth}p{0.7\linewidth}}
\toprule
Task & Zero-shot Prompt \\
\toprule
Selected Juror Names & Can you give me the names of the final list of jurors that were selected to serve on this case as a comma-separated list? Include alternate jurors, if present. Do not output any other text, explanations, or annotations, and make sure to give the juror names.  \\
\midrule
Jury Gender Composition & Count the number of female and male jurors that were chosen to serve on this case. Female jurors are denoted using Ms. in the transcript, and males using Mr. Only count jurors that were chosen to serve on the jury or serve as alternates. Only output the number as a comma-separated list. There should only be 12-14 jurors in total. Do not output any other text, explanations, or annotations. \\
\midrule
Batson Challenges & Can you output if there was a Batson challenge claim made by the defense and state respectively? A Batson challenge happens when a party objects the opposing party's peremptory challenge on grounds that it was used to exclude a potential juror based on race, ethnicity, or sex. Output as a comma-separated value with `Yes' or `No' for each of the parties. Do not output any text, explanations, or annotations. \\
\midrule
Zip Code & Can you give me the zip code of the property in this case? Do not output any other text, explanations, or annotations.\\
\midrule
Landlord Type & Can you give me the type of landlord involved in this case? Output either `Corporation,' `Individual,' or `Boston Housing Authority.' Do not output any other text, explanations, or annotations.\\
\midrule
Landlord Representation & Can you output if the landlord was represented by a counsel. Representation by counsel is if an attorney signed a pleading or court document on behalf of the landlord/tenant at any point during the case. Do not output any other text, explanations, or annotations. If the information is not there, output `N/A'.\\
\midrule
Case Disposition &  Can you give me the ultimate case disposition of this case? Output `VD' for voluntary dismissal by the landlord or both parties, as noted by a Notice of Voluntary Dismissal, a joint Stipulation of Dismissal, or if neither party appeared at court. Output `Default' for a default judgment, noted by a Judgment of Summary Process By Default or a `Default Judgment' entered on the case docket sheet. Output `Dismissed' if the case was dismissed by court order, noted by a court's Notice of Dismissal, if the plaintiff failed to appear at court, or if the case was dismissed by a judge by other means. Output `Trial' if the case underwent trial by a judge. Output `Settlement' if a settlement agreement was reached, noted by a Summary Process Agreement for Judgment. Only output the final case disposition, by chronological ordering. Do not output the initial judgments. Do not output any other text, explanations, or annotations.\\
\midrule
Execution Issued \& \newline Settlement Type & Can you tell me if this case resulted in the issuance of an execution? An execution happens in four ways. First, if the case results in a default judgment, noted by a Judgment of Summary Process By Default or a `Default Judgment' entered on the case docket sheet, an execution must be issued afterwards. Second, if the case results in a trial that awards the landlord a possessory judgment, an execution must be issued afterwards. If the case results in a settlement agreement, noted by a Summary Process Agreement for Judgment, there are two possible pathways. Either the case results in a move-out, where the tenant agrees to vacate the unit on or before a specific date, followed by an execution. Alternatively, the case results in a civil probationary agreement (CPA), containing terms providing for reinstatement of tenancy if the tenant complies with specified terms for a specified period of time, and if a violation of those terms occurs, a motion to issue an execution is filed and granted. In all cases, an issued execution is noted by an Execution of Judgment for Summary Process or an `X Prepared' entered on the docket sheet. If a motion to issue is filed, a granted motion may also include comments such as `Motion Allowed' on the document. Output the settlement type (``Move-out'' or ``CPA'') and whether or not an execution was issued (``Yes'' or ``No'') as a comma-separated list. In settlement cases with multiple settlements, only look at the terms of the initial settlement. If the case is not a default, trial, or settlement case, output `N/A' for both values. Do not output any other text, explanations, or annotations.\\
\bottomrule
\end{tabular} 
\caption{Zero-Shot Prompts}
\label{tab:prompts}
\end{table*}

\clearpage
\clearpage
\bibliography{refs}

\onecolumn
\section{Appendix}
\begin{figure}[!h]
    \centering
    \captionsetup{justification=centering}
     \begin{subfigure}[b]{0.3\textheight}
         \centering
         \includegraphics[width=\textwidth]{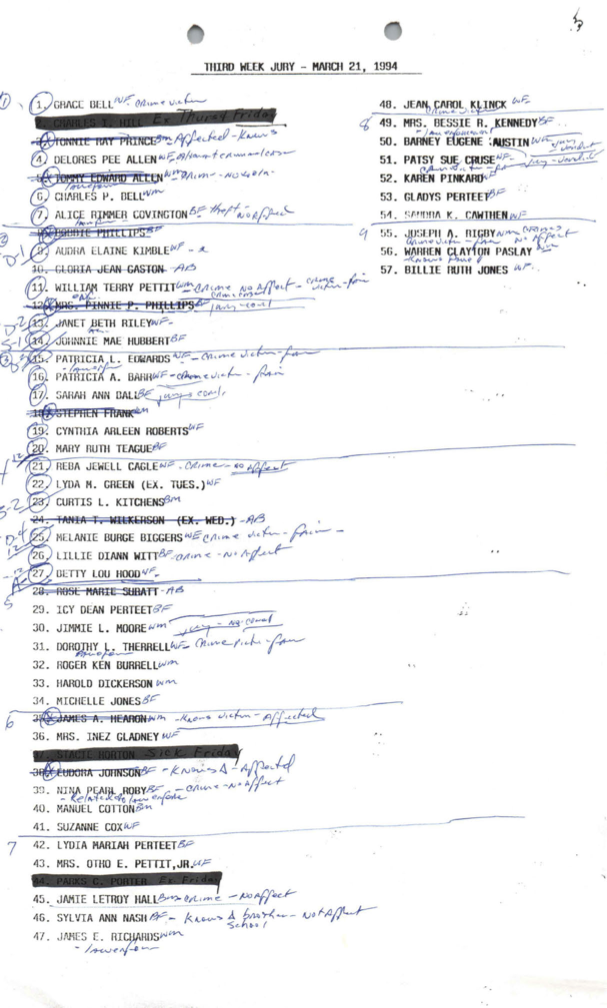}
     \end{subfigure}
     \begin{subfigure}[b]{0.35\textheight}
         \centering
         \includegraphics[width=\textwidth]{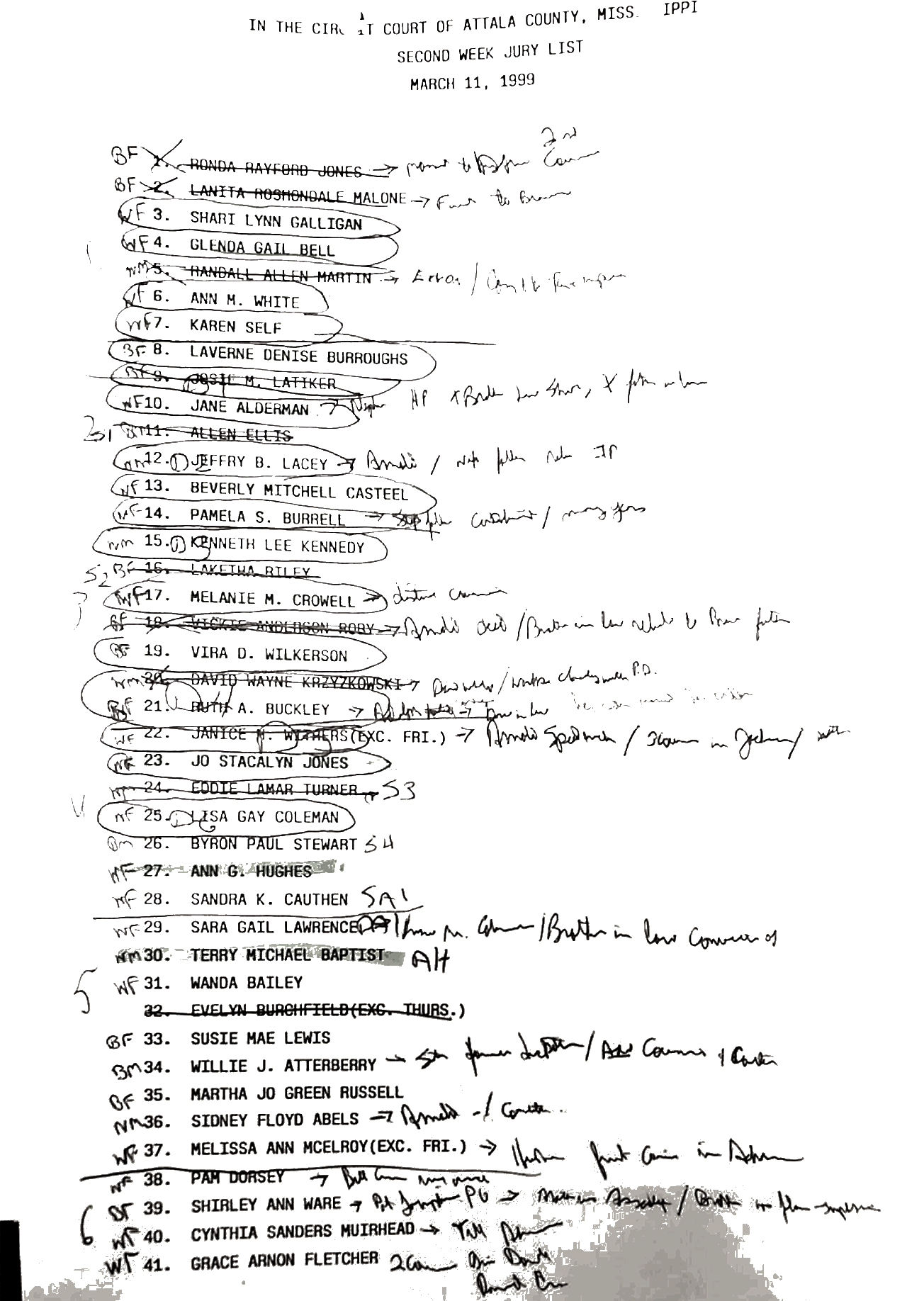}
     \end{subfigure}
     \hfill\\
     \begin{subfigure}[c]{0.3\textheight}
         \centering
         \includegraphics[width=\textwidth]{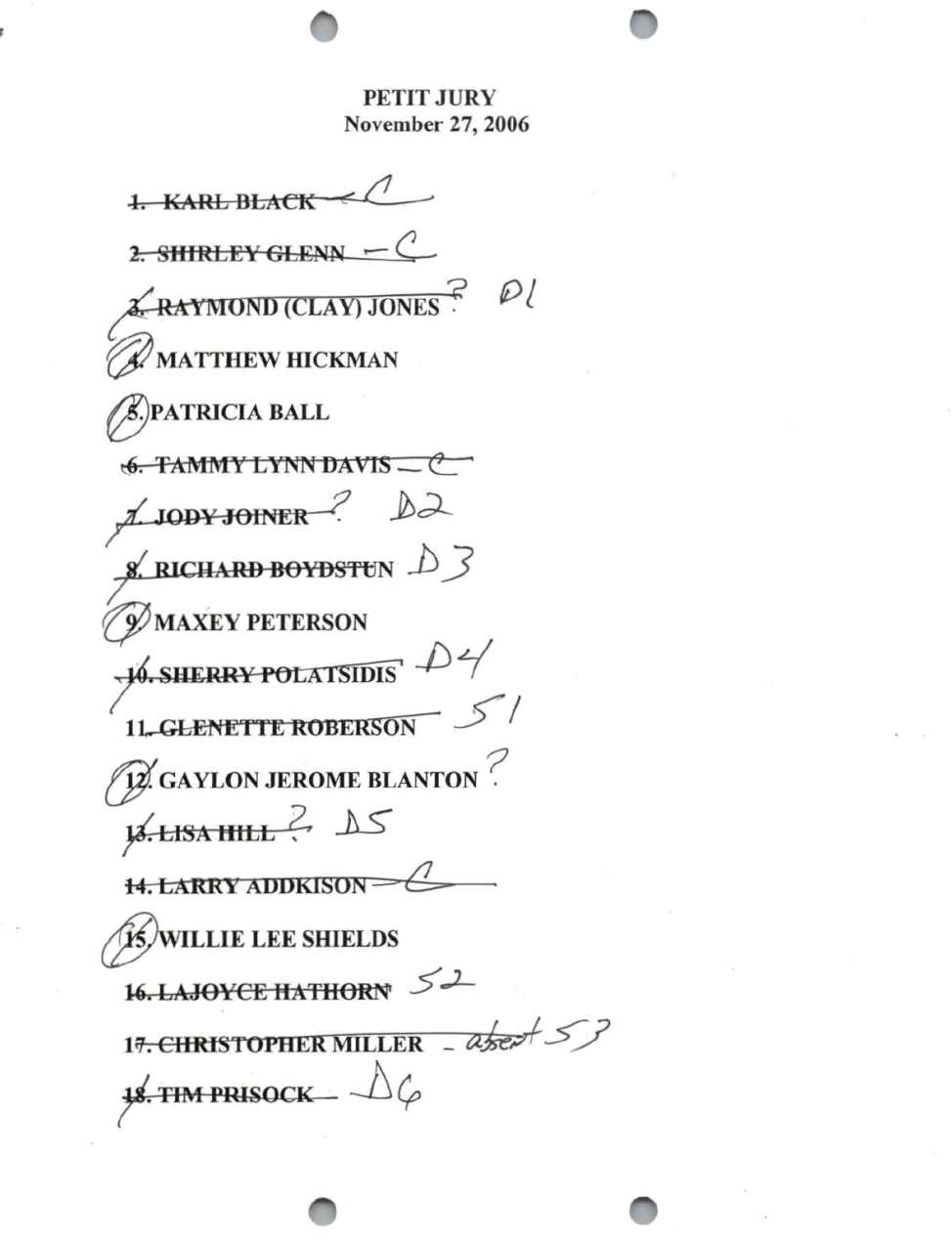}
     \end{subfigure}
     \begin{subfigure}[c]{0.5\textwidth}
         \centering
         \includegraphics[width=\textwidth]{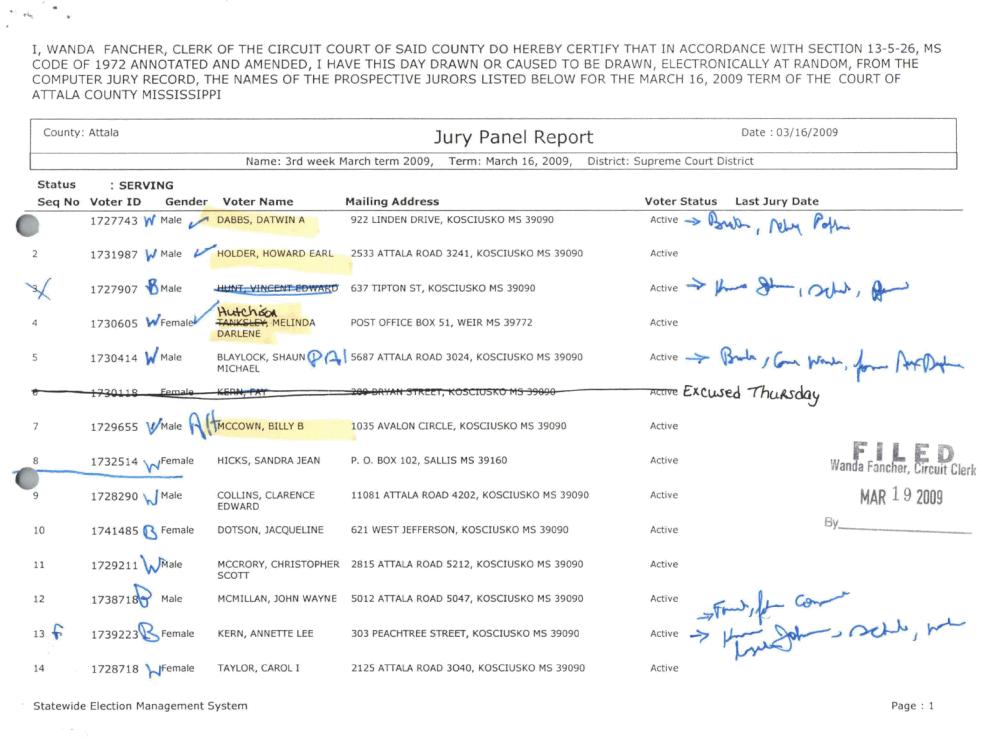}
     \end{subfigure}
    \caption{Example strike sheets showing the variance in note-taking that occurs to document juror demographics and strike status. Common demarcations include 'W'/'B' for race, 'F'/'M' for gender, SX/DX for state and defense strikes, and 'C' for for-cause strikes.}
    \label{fig:example_strike_sheets}
\end{figure}
\begin{figure}[!h]
    \centering
    \captionsetup{width=.8\linewidth}    \captionsetup{justification=centering}
    \includegraphics[width=\textwidth]{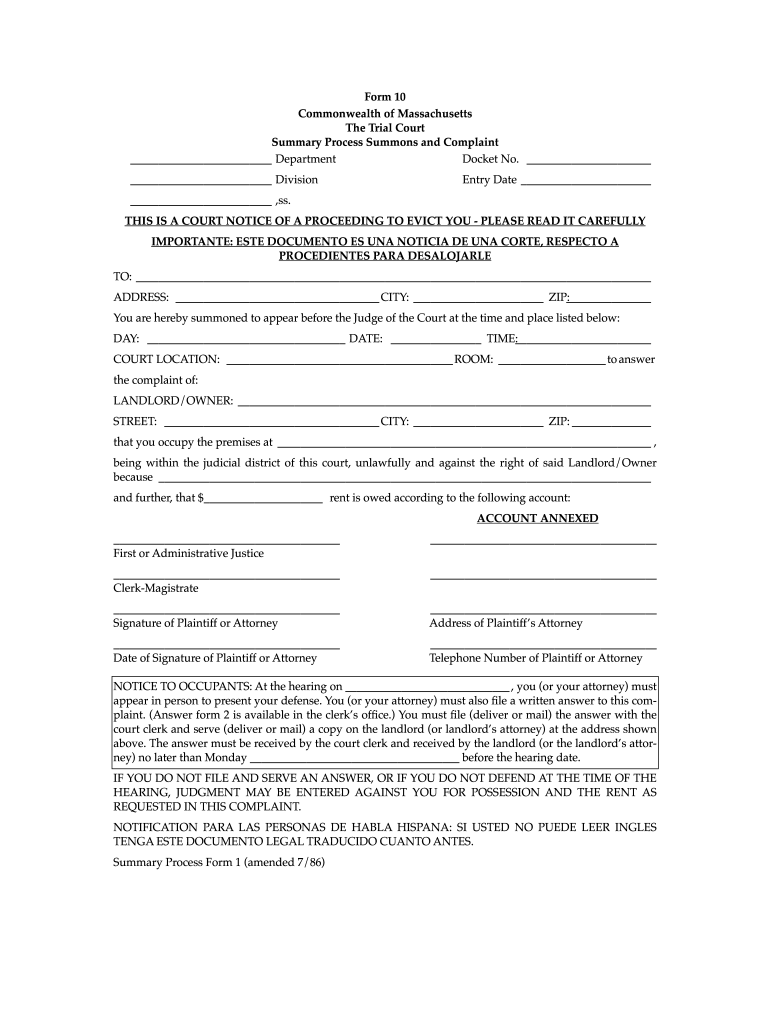}
    \caption{Example Summary Process Summons and Complaint issued by the landlord to call the tenant to court and inform them of the grounds of eviction.}
    \label{fig:example-sc}
\end{figure}
\centering
\begin{figure}
    \centering
    \captionsetup{justification=centering}
    \captionsetup{width=.8\linewidth}
    \includegraphics[width=0.8\textwidth]{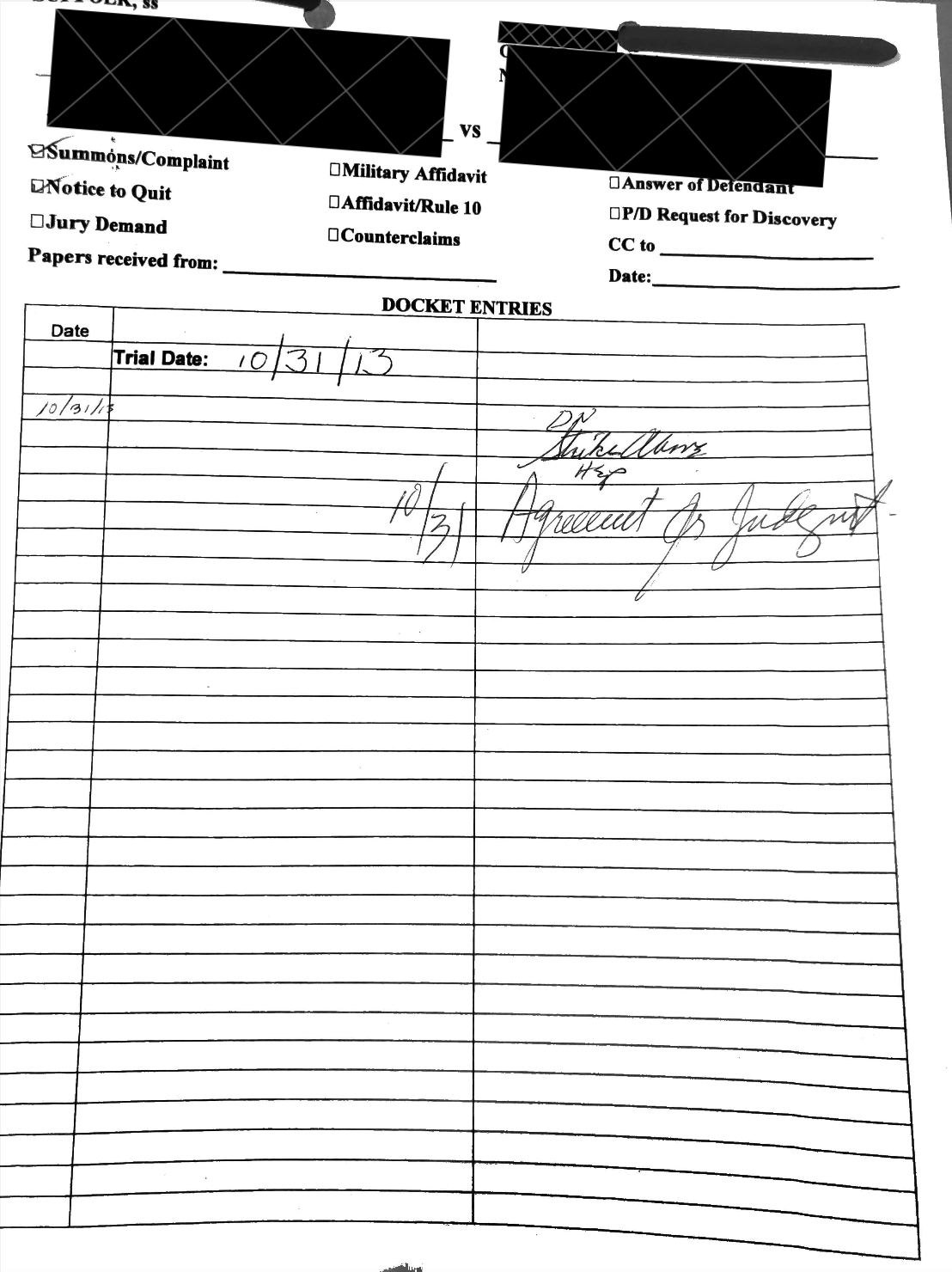}
    \caption{Example docket entry page including the final disposition (Agreement for Judgement) of an eviction case. The variability in handwriting and format of this page makes it difficult to automatically extract information.}
    \label{fig:example-docket}
\end{figure}
\centering

\begin{table*}[!h]
\centering
\begin{tabular}{ccc|ccc} 
\toprule
Context Type & Query Type & \# Cases & Precision & Recall & F1\\
\toprule
Full Transcript & Zero Shot & 50 & 0.82 $\pm$ 0.037 & 0.82 $\pm$ 0.037  & 0.82\\
\midrule
Full Transcript & Two Shot & 50 & 0.95 $\pm$ 0.014 & 0.95 $\pm$ 0.014  & 0.95\\
\midrule
Excerpt & Zero Shot & 50 & \textbf{1.0 $\pm$ 0.0} & \textbf{1.0 $\pm$ 0.0}  & \textbf{1.0}\\
\bottomrule
\end{tabular} 
\caption{Average metrics for the jury selection task of name extraction by context type, query type, and the number of cases across 5 iterations per case. Reported with standard error across the number of cases.}
\label{tab:easy-full}
\end{table*}

\begin{table*}[!h]
\centering
\begin{tabular}{ccc|cccc} 
\toprule
Context Type & Query Type & \# Cases & Female Absolute Error & Male Absolute Error & Total Absolute Error & Accuracy\\
\toprule
Full Transcript & Zero Shot & 50 & 2.07 $\pm$ 0.172 & 2.01 $\pm$ 0.171 & 4.09 $\pm$ 0.305 & 0.036\\
\midrule
Full Transcript & Two Shot & 50 & 1.28 $\pm$ 0.134 & 1.33 $\pm$ 0.141 & 2.61 $\pm$ 0.263 &0.184\\
\midrule
Excerpt & Zero Shot & 50 & 1.08 $\pm$ 0.114 & 1.09 $\pm$ 0.133  & 2.17 $\pm$ 0.233 & 0.238\\
\midrule
Excerpt & Fine-Tuned & 50 & \textbf{0.63 $\pm$ 0.028} & \textbf{0.77 $\pm$ 0.04}  & \textbf{1.40 $\pm$ 0.053} & \textbf{0.340}\\
\bottomrule
\end{tabular} 
\caption{Average metrics for the jury selection task of gender aggregation by context type, query type, and the number of cases across 5 iterations per case. Reported with standard error across the number of cases}
\label{tab:medium-full}
\end{table*}

\begin{table*}[!h]
\centering
\begin{tabular}{ccc|ccc} 
\toprule
Context Type & Query Type & \# Cases & Defense Accuracy & State Accuracy & Accuracy\\
\toprule
Full Transcript & Zero Shot & 75 & 0.669 $\pm$ 0.035 & 0.464 $\pm$ 0.050  & 0.235\\
\midrule
Full Transcript & Two Shot & 75 & \textbf{0.837 $\pm$ 0.026} & \textbf{0.872 $\pm$ 0.012} & \textbf{0.755}\\
\midrule
\midrule
Full Transcript & Zero Shot & 50 & 0.684 $\pm$ 0.045 & 0.460 $\pm$ 0.062  & 0.232\\
\midrule
Full Transcript & Two Shot & 50 & \textbf{0.848 $\pm$ 0.032} & \textbf{0.892 $\pm$ 0.02}  & \textbf{0.768}\\
\bottomrule
\end{tabular} 
\caption{Average metrics for the jury selection task of identifying Batson challenges, by context type, query type, and the number of cases across 5 iterations per case. Reported with standard error across the number of cases.}
\label{tab:hard-full}
\end{table*}
\begin{table*}[t!]
\centering
\captionsetup{justification=centering}
\begin{tabular}{lp{2.5cm}} 
\toprule
\textbf{County} & \textbf{Original \newline F:M Ratio}\\
\toprule
Winston & 2.415 \\
\midrule
Attala & 2.250 \\
\midrule
Montgomery & 2.216\\
\midrule
Grenada & 1.586 \\
\midrule
Carroll & 1.578 \\
\midrule
Choctaw & 0.947 \\
\midrule
Webster & 0.741 \\
\bottomrule
\end{tabular} 
\quad
\begin{tabular}{lp{2.5cm}} 
\toprule
\textbf{County} & \textbf{Baseline LLM\newline F:M Ratio}\\
\toprule
Montgomery $\uparrow$ & 1.596 \\
\midrule
Attala & 1.491  \\
\midrule
Winston $\downarrow$ & 1.422 \\
\midrule
Carroll $\uparrow$ & 1.330 \\
\midrule
Grenada $\downarrow$ & 1.272 \\
\midrule
Choctaw & 1.094 \\
\midrule
Webster & 0.741 \\
\bottomrule
\end{tabular} 
\quad
\begin{tabular}{lp{2.7cm}} 
\toprule
\textbf{County} & \textbf{Fine-Tuned LLM\newline F:M Ratio}\\
\toprule
Winston & 1.752 \\
\midrule
Montgomery $\uparrow$ & 1.663\\
\midrule
Attala $\downarrow$ & 1.638 \\
\midrule
Grenada & 1.423 \\
\midrule
Carroll & 1.333 \\
\midrule
Choctaw & 0.971 \\
\midrule
Webster & 0.741 \\
\bottomrule
\end{tabular} 
\caption{Original and inferred rankings of counties based on the average female-to-male ratio of jury composition. Shown for both the baseline and fine-tuned LLMs.}
\label{tab:gender-county}
\end{table*}

\begin{table*}[t!]
\centering
\begin{tabular}{lp{2cm}} 
\toprule
\textbf{Prosecutor} & \textbf{Original \newline F:M Ratio}\\
\toprule
Doug Evans & 2.374 \\
\midrule
Susan Denley & 2.088 \\
\midrule
Walter Bleck & 1.930 \\
\midrule
Mickey Mallette & 1.816\\
\midrule
Michael Howie & 1.710 \\
\midrule
Chet Kirkham & 1.400 \\
\midrule
Clyde Hill & 1.320 \\
\midrule
Greg Meyer & 1.228 \\
\midrule
Kevin Horan & 1.118 \\
\midrule
Ryan Berry & 1.113 \\
\bottomrule
\end{tabular} 
\quad
\begin{tabular}{lp{2.25cm}} 
\toprule
\textbf{Prosecutor} & \textbf{Baseline LLM \newline F:M Ratio}\\
\toprule
Susan Denley $\uparrow$ & 1.654 \\
\midrule
Walter Bleck $\uparrow$& 1.569 \\
\midrule
Mickey Mallette $\uparrow$ & 1.472 \\
\midrule
Doug Evans $\downarrow$ & 1.439  \\
\midrule
Michael Howie & 1.388 \\
\midrule
Kevin Horan $\uparrow$ & 1.216 \\
\midrule
Greg Meyer $\uparrow$ & 1.167 \\
\midrule
Clyde Hill $\downarrow$ & 1.074 \\
\midrule
Ryan Berry $\uparrow$ & 1.04 \\
\midrule
Chet Kirkham $\downarrow$ & 0.75 \\
\bottomrule
\end{tabular} 
\quad
\begin{tabular}{lp{2.6cm}} 
\toprule
\textbf{Prosecutor} & \textbf{Fine-Tuned LLM\newline F:M Ratio}\\
\toprule
Doug Evans & 1.699 \\
\midrule
Susan Denley & 1.654 \\
\midrule
Walter Bleck & 1.619 \\
\midrule
Chet Kirkham $\uparrow$ & 1.6 \\
\midrule
Mickey Mallette $\downarrow$ & 1.426\\
\midrule
Michael Howie $\downarrow$ & 1.404 \\
\midrule
Greg Meyer $\uparrow$ & 1.228 \\
\midrule
Clyde Hill $\downarrow$ & 1.193 \\
\midrule
Kevin Horan  & 1.144 \\
\midrule
Ryan Berry & 1.012 \\
\bottomrule
\end{tabular} 
\caption{\centering Original and inferred rankings of prosectors based on the average female-to-male ratio of jury composition. Shown for both the baseline and fine-tuned LLMs.}
\label{tab:gender-pros}
\end{table*}

\end{document}